\documentclass[
reprint,
aps,
prx
]{revtex4-2}

\usepackage{natbib}
\usepackage{graphicx}
\usepackage{dcolumn}
\usepackage{bm}
\usepackage{csquotes}
\usepackage{siunitx}
\usepackage{IEEEtrantools}
\usepackage{braket}
\usepackage{amsmath}
\usepackage{amssymb}
\usepackage{nicefrac}
\usepackage{xcolor}
\usepackage{verbatim}
\usepackage{hyperref}
\hypersetup{
    colorlinks,
    citecolor=blue,
    filecolor=blue,
    linkcolor=black,
    urlcolor=blue
}

\DeclareSIUnit\gauss{G}
\newcommand{\rhoss}{\rho_{ee}^{\text{st}}}
\newcommand{\deltam}{\delta'_{\mathrm{m}}}
\newcommand{\deltals}{\delta_{\text{LS}}}

\newcommand{\Omegat}{\tilde{\Omega}}
\newcommand{\Deltat}{\tilde{\Delta}}
\newcommand{\deltac}{\delta_\text{c}}

\begin{document}

\title{Raman imaging of atoms inside a high-bandwidth cavity}

\author{E. Uru\~{n}uela}
\email{Corresponding author: e.urunuela@iap.uni-bonn.de}
\author{M. Ammenwerth}
\author{P. Malik}
\author{L. Ahlheit}
\author{H. Pfeifer}
\author{W. Alt}
\author{D. Meschede}
\affiliation{Institute for Applied Physics, University of Bonn, Wegelerstr. 8, 53115 Bonn, Germany}

\date{\today}

\begin{abstract}
    High-bandwidth, fiber-based optical cavities are a promising building block for future quantum networks. They are used to resonantly couple stationary qubits such as single or multiple atoms with photons routing quantum information into a fiber network at high rates. In high-bandwidth cavities, standard fluorescence imaging on the atom-cavity resonance line for controlling atom positions is impaired since the Purcell effect strongly suppresses all-directional fluorescence. Here, we restore imaging of $^{87}$Rb atoms strongly coupled to such a fiber Fabry-Pérot cavity by detecting the repumper fluorescence which is generated by continuous and three-dimensional Raman sideband cooling. We have carried out a detailed spectroscopic investigation of the repumper-induced differential light shifts affecting the Raman resonance, dependent on intensity and detuning. Our analysis identifies a compromise regime between imaging signal-to-noise ratio and survival rate, where physical insight into the role of dipole-force fluctuations in the heating dynamics of trapped atoms is gained.
\end{abstract}

\maketitle

\section{Introduction}
\label{sec:introduction} 

\begin{figure}[t]
    \centering
    \includegraphics{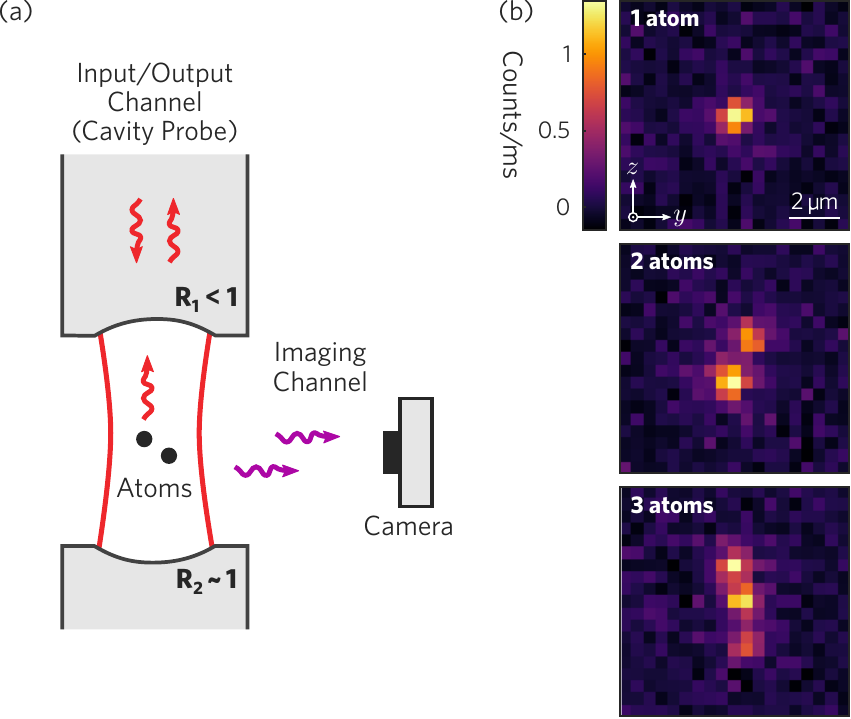}
    \caption{
    (a) A fiber based, single-sided high-bandwidth optical cavity (reflectivities $R_1< R_2 \simeq 1$) serves to efficiently interface atomic qubits with photonic quantum information routed on a fiber network.
    (b) Images of small atomic ensembles (1, 2 and 3 atoms) at the center of our high-bandwidth cavity registered with repumper fluorescence during continuous Raman sideband cooling in 3D (exposure time \SI{1}{\second}).} 
    \label{fig:idea}
\end{figure}

A major goal pursued in quantum communication science during the last decades has been to create quantum networks \cite{Gisin2007,VanMeter2014} that would allow to distribute quantum information in an analogous way to classical information with the present internet \cite{Kimble2008,Wehner2018}. In such proposed networks, flying qubits (i.e. photons) are excellent carriers of quantum information in fiber-based networks connecting quantum nodes capable of generating, processing and storing quantum information in the form of stationary matter qubits \cite{Duan2001}. It has been demonstrated  \cite{vanEnk2004,Ritter2012,Reiserer2015} that optical cavities can enhance light matter interaction to the level required for efficient interfacing of incoming and outgoing photonic information at the nodes of future quantum networks. 

Much research effort is still invested in finding the best suited physical platform for the nodes \cite{Duan2010,Sangouard2011,Awschalom2021}, having to fulfill challenging technical requirements to enable efficient and coherent exchange of quantum information between the photons and the matter counterpart.  Several different physical platforms are considered for the realization of such nodes \cite{vanLoock2020}, of which atoms stored in and strongly coupled to optical cavities offer a good compromise of efficient light matter interaction and long storage times of quantum information \cite{Cirac1997,Kimble1998,Miller2005,Haroche2006}. Hence, fiber-based optical cavities \cite{Hunger2010,Gallego2016,Pfeifer2022} coupling to atomic matter qubits make an attractive solution for directly routing quantum information via fiber links.

A promising regime of fiber-based Fabry-Pérot cavities (FFPCs) is to realize strong atom-field coupling $g$ along with a cavity of high-bandwidth $\kappa$ \cite{Gallego2018}, characterized by $g, \kappa \geq \gamma$ and $g\approx\kappa$, with $\gamma$ the free space decay rate of the atom. The main breakthrough of such miniaturized cavities is that they allow for information flow at high rates of order $\kappa$, while still providing the necessary conditions for the strong coupling regime with cooperativity $C=\frac{g^2}{2\kappa\gamma}\gg1$. This unique feature of FFPCs is possible due to their small mode volume $V$ that boosts the coupling strength $g \propto \frac{1}{\sqrt{V}}$. The single-sided configuration of Fig.~\ref{fig:idea}(a) further positions FFPCs as an ideal platform for quantum nodes: the intrinsic fiber coupling to a single input-output channel enables convenient and efficient routing of photons carrying quantum information from and to the optical fiber, i.e. the quantum channel, mediated by the cavity-coupled atoms. 

In recent years, important basic functionalities of quantum nodes have been demonstrated in proof-of-principle experiments, with a single atom (or few atoms) in FFPCs and in bulk cavities. These advances include highly efficient and deterministic single photon sources \cite{Kuhn2010,NisbetJones2011}, the storage and retrieval of a single photon into a single atom \cite{Wilk2007} even beyond the adiabatic regime \cite{Macha2020}, the bandwidth conversion of a single photon mediated by a single cavity-coupled atom \cite{Morin2019}, and recently prototypes for quantum memories and quantum repeaters~\cite{Langenfeld2021}.

The functionality of the atom-cavity system can be extended by increasing the number of atoms \cite{Reitz2022} to overcome the limit for the single-atom coupling strength ($g$ is bounded by the smallest mode volume $V$ technically achievable), by means of the Dicke enhancement with $N$ identically coupled atoms, yielding $g_N \propto \sqrt{N}\cdot g$. The Dicke enhanced rate $g_N$ \cite{Dicke1954,Fleischhauer2000,Gorshkov2007,Colombe2007} can boost the coherent photon-atom ensemble interaction to the level that Purcell effect allows to match the bandwidths of very diverse quantum emitters, e.g. semiconductor quantum dots ($\gamma \approx \SI{1}{\giga\hertz}$) and neutral atoms ($\gamma \approx \SI{6}{\mega\hertz}$). Ensembles of atoms in cavities have also been envisaged as promising platforms for quantum simulators~\cite{Altman2021}.

The necessary level of control for both single and multi-atom implementations requires optical tweezers or lattices for trapping, and motional control of the atoms \cite{Bloch2005, Meschede2006, Ott2016}. The atom-cavity coupling strength is governed by the local field strengths and hence knowledge of the atoms positions within the cavity is essential for operating efficient protocols, e.g. for photon storage \cite{Gorshkov2007} or dissipative entanglement \cite{Kastoryano2011}. In the case of the FFPC setup shown in Fig.~\ref{fig:idea}(a) this knowledge is extracted through a secondary imaging channel, independently of the primary fiber quantum channel. Fluorescence imaging, complemented by cooling measures to compensate scattering-induced heating and atom loss, has been established for long as the workhorse for positional detection of individual neutral atoms \cite{Alberti2016,Martinez-Dorantes2018}.

High-bandwidth cavities, however, are impairing the standard methods of both cooling and imaging since the strong Purcell effect causes almost all light emitted by the atoms at the cavity resonance line to be routed into the fiber channel, which amounts to suppressing fluorescence in the direction transverse to the cavity axis \cite{Gallego2018}. Also, the convenient cavity cooling technique \cite{Ritsch2013} does not exist for $\kappa \gg \gamma$, it is switched off by the large cavity decay. 

Alternative cooling schemes compatible with high bandwidth cavities and restricted FFPC geometries have been established previously using Raman sideband cooling techniques \cite{Reimann2014,Neuzner2018}. Here we show that the detection of the repumper fluorescence (which is not subject to the Purcell effect) emitted during the Raman cooling cycles, provides enough intensity to allow imaging of the atoms trapped in a 3D lattice superposed with the cavity field.

The technique dubbed Raman imaging was pioneered by \cite{Lester2014,Patil2014} and quickly adopted in the field of quantum gas microscopes \cite{Cheuk2015,Parsons2015,Omran2015} as a powerful method for single-site resolved and loss-free imaging of dense atomic ensembles in optical lattices \cite{Ott2016}. In our experiment, we customize this method to the FFPC setup in Fig.~\ref{fig:idea}(a), where the restricted optical access renders coupling of one of the Raman beams and the repumper copropagating with the fiber quantum channel advantageous. The detrimental Purcell effect is avoided by choosing atomic transitions for the imaging channel with large detuning from the quantum channel (resonant with the cavity). This implementation allows us to obtain clear pictures of the trapped atoms with \SI{1}{\second} exposure time, as shown in Fig.~\ref{fig:idea}(b), while simultaneously probing the atom-cavity coupling through the reserved quantum channel.

Combining the information obtained from both the imaging and the fiber quantum channel not only allows us to optimize the imaging scheme, but also to perform an in depth analysis of the underlying physical effects. We study the repumper-induced differential light shifts that occur during Raman sideband cooling and analyze the complex parameter space governing both the scattering rate and the balance between cooling efficiency and heating rates. Our analysis provides useful parameter regions for a compromise between high survival rate and an acceptable imaging signal-to-noise ratio. Additionally, we gain physical insight into the underlying heating dynamics for atoms in optical lattices under near-resonant illumination, originating from dipole-force fluctuations \cite{Taieb1994,Martinez-Dorantes2018}. For certain parameters we observe high heating rates which we discuss and validate with a semi-classical Monte Carlo simulation of scattering dynamics in dressed-state potentials.

\section{Experimental Setup and Methods}
\label{sec:setup}

\begin{figure}[t]
    \centering
    \includegraphics{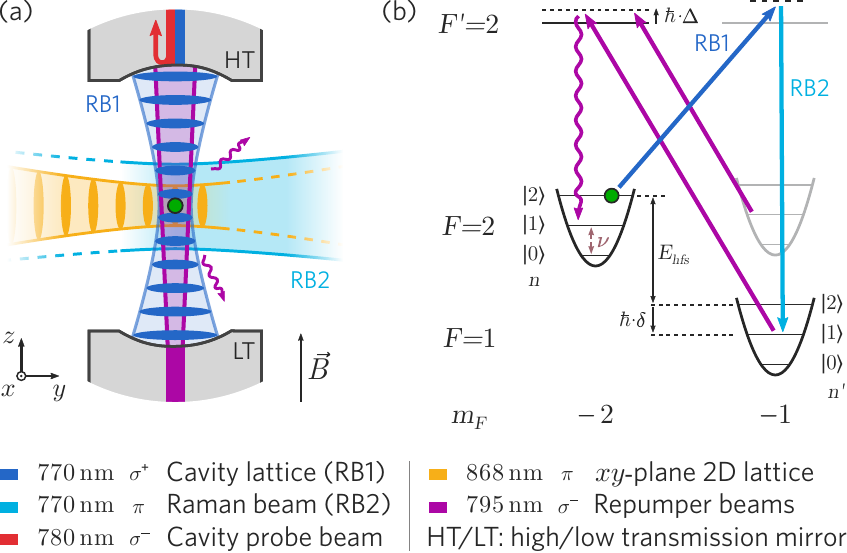}
    \caption{
    Experimental setup and employed transitions for imaging of $^{87}$Rb atoms inside a high-bandwidth cavity. 
    (a) Atoms are confined in the $xy$ plane by a red-detuned 2D-lattice  ($x$ direction not shown), and along the cavity by a blue-detuned lattice. The \SI{770}{\nano\metre} intra-cavity field also acts as a first Raman beam (RB1), and the second one (RB2) is overlapped with the $y$-axis lattice. The repumper beams induce scattering of photons on the D1 line that are collected to image the atoms onto an EMCCD camera. We probe the cavity on the D2 line to detect the presence and internal state of the atom. 
    (b) Representation of the 1D cooling scheme. The frequency difference of RB1 and RB2 equals the Zeeman-shifted energy splitting between the involved hyperfine levels $E_{\textit{hfs}}$, plus a two-photon detuning $\delta$. To drive Raman transitions on the cooling sideband, $\delta$ is tuned to the near-harmonic trap frequency $\nu$. To close the cooling cycle and obtain images with high SNR, the repumper beams are kept detuned from the D1 line by $\Delta$.} 
    \label{fig:setup}
\end{figure}

In our experiment, few $^{87}$Rb atoms are trapped in a 3D lattice and strongly coupled to a high-bandwidth fiber Fabry-Pérot cavity (FFPC) with parameters $(g, \kappa, \gamma)=2\pi \cdot (80, 41, 3)\,\si{\mega\hertz}$ \cite{Gallego2016,Gallego2018}. A simplified diagram of the experimental setup is shown in Fig.~\ref{fig:setup}(a).

To obtain a directional single-sided resonator, a mirror with high transmission (HT) is used at the input-ouput-channel side. A \SI{780}{\nano\metre} cavity probe beam is coupled through this port to resonantly interrogate the atoms. The cavity is stabilized to a length that features a simultaneous resonance with the probe light at the D2 line of $^{87}$Rb, and with \SI{770}{\nano\metre} light forming a blue-detuned intra-cavity optical lattice (DT$_z$). In this configuration, trapping sites at intensity minima of DT$_z$ coincide with intensity maxima of the probe standing wave at the cavity center, providing an atom-cavity interaction in the strong coupling regime~\cite{Gallego2018}. 

The atoms are trapped in a 3D lattice at the center of the cavity. In the $xy$ plane, the lattice is formed by two near-orthogonal red-detuned \SI{868}{\nano\metre} standing-wave dipole traps (DT$_x$ and DT$_y$), and in the $z$ axis by the blue-detuned \SI{770}{\nano\metre} intra-cavity lattice (DT$_z$). With a depth of $\sim\!\SI{0.5}{\milli\kelvin}$ in each direction this allows for trapping in the Lamb-Dicke regime \cite{Leibfried2003}, in a region defined by the waist of the beams $(w_{x}, w_{y},w_{z})=(13, 11, 5)\,\si{\micro\meter}$. The lattice polarizations are defined as $(P_{x}, P_{y},P_{z})=(\pi,\pi,\sigma^+)$, with the quantization axis of our system set parallel to the cavity axis by applying a magnetic guiding field of $\sim\!\SI{1.8}{\gauss}$. The dipole trap along the $y$ axis serves as an optical conveyor belt \cite{Schrader2001,Kuhr2001} to transport single atoms to the cavity center. The conveyor belt is loaded from a magneto-optical trap (MOT) \SI{1}{\milli\metre} away.

For Raman cooling we use two-photon transitions that couple the internal and external degrees of freedom of the atom. The intra-cavity optical lattice DT$_z$ also plays the role of the first Raman beam (RB1). The second \SI{770}{\nano\metre} Raman beam (RB2) is phase locked to DT$_z$ and co-propagates along DT$_y$ enabling momentum transfer on each cooling cycle. For long term frequency stability, the source for RB2 is a distributed Bragg reflector (DBR) laser. It is upgraded with an external optical feedback to reduce its linewidth and enable phase locking (see Appendix \ref{sec:Raman-LRDBR}). To achieve Raman coupling for all three lattice directions, the lattice beams DT$_{x,y}$ are not fully orthogonal to DT$_z$, but feature small angles with respect to the normal plane of the vertical lattice ($\theta_x\approx\ang{15}$ and $\theta_y\approx\ang{6}$). This geometry ensures that the difference of the Raman beam wave vectors has a projection along all lattice dimensions. With uninterrupted cooling we observe vacuum-limited trapping $1/e$ lifetimes of $\sim\SI{1}{\minute}$ for single atoms in the lattice.

To close the cooling cycles we couple two $\sigma^-$-polarized \SI{795}{\nano\metre} optical pumping beams through the lower transmission cavity mirror (LT). One of them is used to pump the atomic population from the lower to the higher hyperfine ground state, while the other is used to polarize the atoms in $m_F = -2$. For the rest of the paper we refer to them together as repumper beams (in plural), and in singular to the hyperfine-changing repumper alone. These beams also provide illumination light for fluorescence imaging of the atoms. They are tuned to the atomic D1 line, hence they provide an imaging channel that is fully independent from the cavity interaction on the D2 line, and thus insensitive to the Purcell effect.

In order to obtain information about the atoms inside the FFPC we use two schemes. First, a non-destructive cavity probe measurement determines the presence of an atom inside the cavity and its internal state. If an atom couples to the cavity mode, the induced vacuum-Rabi splitting increases the reflection signal of the cavity-resonant probe light, featuring a binary readout of the atom-cavity coupling state \cite{Boozer2006,Gallego2018}. Second, fluorescence imaging provides knowledge on the number of atoms and their position within the cavity mode. For this, photons scattered by the atoms are collected with an in-vacuum high-NA lens along the $x$ axis and recorded with an EMCCD camera. The combination of these two detection techniques allows us to extract complementary and independent information on the atom trapping lifetime and on the photon scattering rate, respectively, which we use for the subsequent analysis.

\section{Imaging with Raman cooling}
\label{sec:methods}

Our imaging technique of atoms inside fast cavities is based on inducing scattering of photons with near-resonant illumination while keeping the atoms close to their motional ground state. This is achieved by applying resolved Raman sideband cooling in all three lattice dimensions, and simultaneous imaging of the photons scattered during the repumping transition onto the EMCCD camera.

To describe this cooling method, we consider the case of a single atom in a one-dimensional lattice in the Lamb-Dicke regime, as shown in Fig.~\ref{fig:setup}(b). In the harmonic approximation for a cold atom at the bottom of the trap, the quantized motional energy levels are given as $E_n = \hbar\nu (n+\nicefrac{1}{2})$, with $n=0,1,2,...$, and $\nu/2\pi$ being the trap oscillation frequency. In the following, we use the notation $\ket{F, m_F ;n}$ to represent the full atomic quantum state, with its ground state $5^2 \text{S}_{\nicefrac{1}{2}}$ hyperfine level denoted by $\ket{F}$, its magnetic sublevel by $\ket{m_F}$  and its vibrational state by $\ket{n}$.

\begin{figure}[t]
    \centering
    \includegraphics{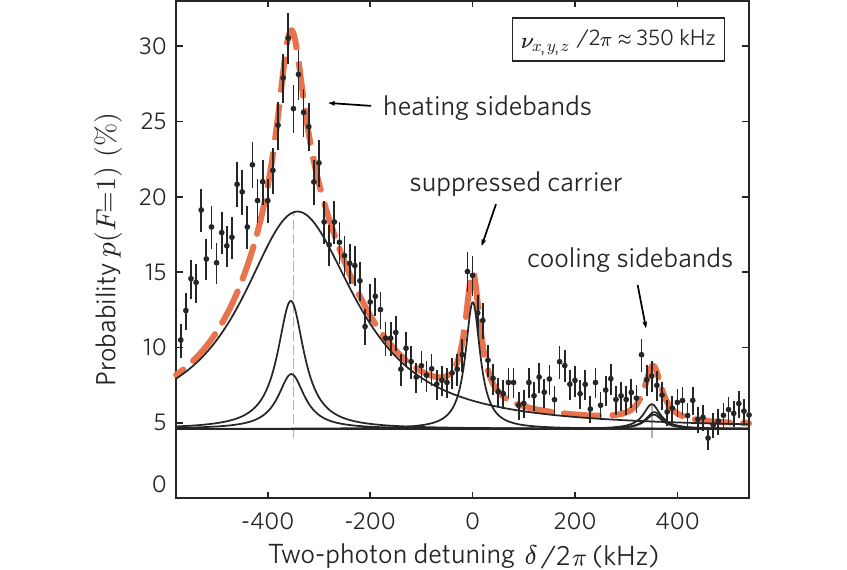}
    \caption{
    Raman spectrum of a single atom in the 3D lattice after \SI{5}{\milli\second} of continuous Raman sideband cooling (cRSC), addressing the overlapped cooling sidebands with a single Raman-beam pair (see main text and Appendix \ref{sec:rsc3d}). The usage of the intra-cavity field as a Raman beam highly suppresses unwanted off-resonant carrier transitions \cite{Reimann2014,Neuzner2018}. We observe the cooling efficacy by a motional ground state population $>\!\SI{85}{\percent}$ in each direction estimated from sideband imbalance, and by single-atom $1/e$ trapping lifetimes of $\sim\!\SI{1}{\minute}$. 
    } \label{fig:simpleSpectrum}
\end{figure}

A cooling cycle starts with the atom optically pumped to the ground state $\ket{2, -2 ;n}$, by means of the repumper beams tuned to the D1 transitions between hyperfine $F$ and $F^\prime$ levels $1\rightarrow 2^\prime$ and $2\rightarrow 2^\prime$ respectively. The Raman beams RB1 and RB2 are phase locked with a frequency difference corresponding to the energy splitting $E_{\textit{hfs}}$ between the outermost hyperfine levels $\ket{2, -2}$ and $\ket{1, -1}$ (Zeeman-shifted by $\SI{-1.27}{\mega\hertz}$), plus a two-photon detuning $\delta$. We note that throughout the paper this detuning is referenced to the carrier transition without light shifts at $\delta=0$. We tune the two-photon Raman resonance on the cooling sideband with $\delta =\nu$, thus driving the transition $\ket{2, -2 ;n}\rightarrow\ket{1, -1 ;n-1}$. A \SI{5}{\tera\hertz} single-photon detuning suppresses off-resonant scattering by the Raman beams. The repumper beams, detuned by $\Delta$ from the excited state $5^2 \text{P}_{\nicefrac{1}{2}} \ket{2',-2}$ , pump the atom from the lower hyperfine state back to $\ket{2, -2 ;n-1}$ effectively reducing its oscillation energy by one quantum. This cycle continues until the ground state of the atomic motion, or till an equilibrium is reached of cooling vs heating processes, due to e.g. photon recoil during repumping. 

To implement Raman sideband cooling in 3D, in contrast to addressing each lattice dimension with independent beams or with multi-tone pulses at the different sideband frequencies \cite{Han2000,Lester2014,Cheuk2015}, we address the cooling transitions along all directions simultaneously with the single Raman-beam pair RB1+RB2. For this, we tune the three trap frequencies to near-degeneracy at $\nu_{x,y,z}/2\pi\approx \SI{350}{\kilo\hertz}$  (for more details see Appendix~\ref{sec:rsc3d}), and we drive the overlapped cooling sidebands with the Raman two-photon detuning  $\delta=\nu_{x,y,z}$. 

A pulsed scheme to drive the cooling transitions, with alternating Raman and repumping pulses \cite{Parsons2015}, is not applicable with the geometry of our beams, since the different directions feature distinct Lamb-Dicke parameters $(\eta_{x},\eta_{y},\eta_{z})=(0.04, 0.12, 0.14)$. This entails mismatched Rabi frequencies on the different (but overlapped) cooling sidebands, and thus different $\pi$-pulse times for each direction, making the optimization difficult. For this reason we have opted for continuous Raman sideband cooling (cRSC), activating the Raman and repumper beams simultaneously with constant intensity during the entire cooling interval.

A Raman spectrum measured after optimization using a cRSC interval of \SI{5}{\milli\second} is displayed in Fig.~\ref{fig:simpleSpectrum}. It shows a large imbalance of all cooling and heating sidebands which gives a clear signature of the high 3D cooling power, with an estimated residual temperature $T\approx\SI{1.4}{\micro\kelvin}$ corresponding to $>\!\SI{85}{\percent}$ motional ground state population in each direction (see App.~\ref{sec:rsc3d}). Note that usage of the intra-cavity field as one of the Raman beams not only solves the problem of optical access, but also suppresses off-resonant coupling on the carrier transition \cite{Reimann2014,Neuzner2018}. 

We obtain fluorescence images of the atoms by collecting the \SI{795}{\nano\meter} photons scattered during the repumping cycle of cRSC, with an in-vacuum 0.5-NA lens, and recording them with the EMCCD camera (Andor iXon 3). The main properties of the imaging system are: collection efficiency $\sim\!\SI{4}{\percent}$, magnification $\times 35$, point spread function width $\sim\!\SI{0.8}{\micro\meter}$ corresponding to \SI{1.72}{px} on the CCD. The imaging fidelity is limited by background scattering from the cavity mirrors that cannot be fully suppressed from the field of view in our miniaturized FFPC setup. This sets a strong condition on employing low repumper intensity. To maintain a sufficiently high scattering rate, this requires the repumper to be near resonant.  An optimal imaging signal-to-noise ratio (SNR)~\cite{Alberti2016} is found for a two-photon Raman Rabi frequency (on the carrier) $\Omegat_0 /2\pi\approx \SI{400}{\kilo\hertz}$, a repumper blue detuning $\sim\!\SI{3}{\mega\hertz}$ and a repumper intensity $\sim\!0.057 I_\text{sat}$ (see Fig.~\ref{fig:RamanImage}). For such parameters we measure a scattering rate per atom of $\sim\!\SI{2e4}{photons\per\second}$ which allows to collect about $\SI{800}{photons}$ during a \SI{1}{s} exposure time (atom  survival probability $> \SI{90}{\percent}$). The fluorescence images shown in Fig.~\ref{fig:idea}(b) have a SNR of $\sim$ 13, which is sufficient to count individual atoms and determine their position with full site resolution~\cite{Alberti2016}.

\section{Differential Light Shifts During Continuous Raman Sideband Cooling}
\label{sec:lightshifts}

\begin{figure*}[t]
    \centering
    \includegraphics{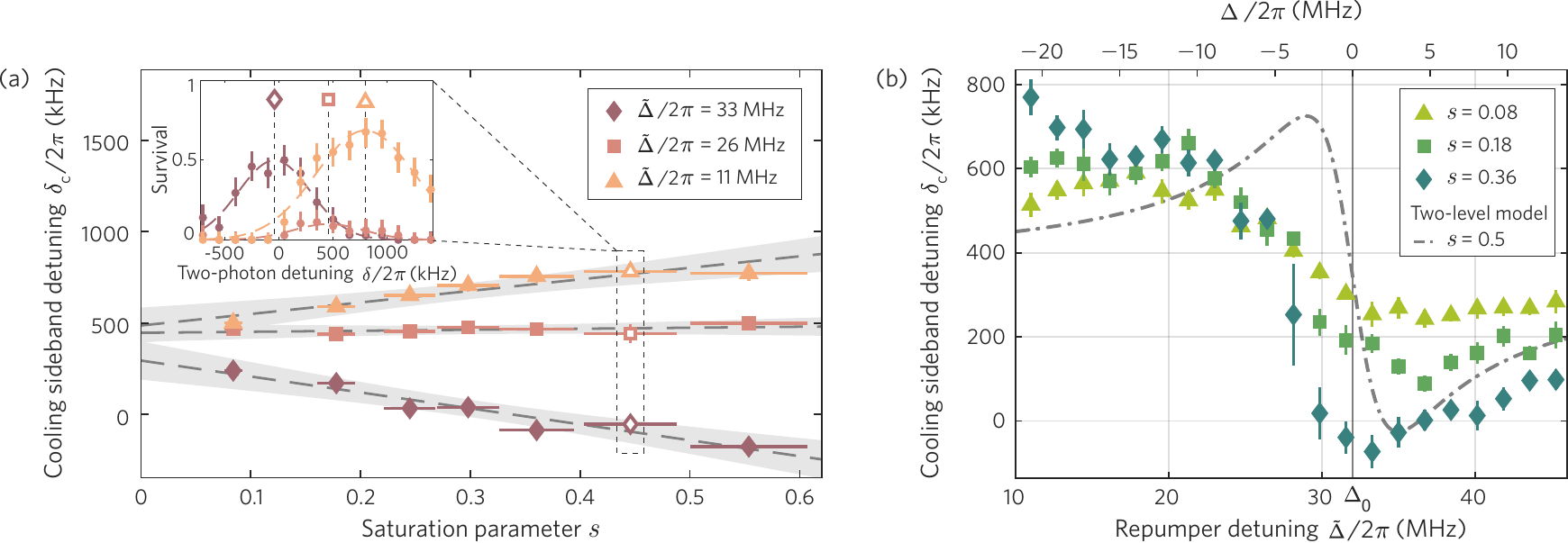}
    \caption{Differential light shifts (i.e. cooling sideband detuning $\deltac=\deltals+\nu$) induced by the near-resonant repumper beams on the hyperfine ground states of $^{87}$Rb. (a)  Dependence on repumper saturation parameter $s=I_{\text{rep}}/I_{\text{sat}}$ for selected detunings~$\Deltat$ shows linear scaling. Inset: For each set of parameters $\{s, \Deltat\}$, the value of $\deltac$ is the Gaussian fit center of the survival probability vs. two-photon detuning~$\delta$. (b) Dispersive shaped light shift dependence on the repumper detuning for selected $s$  values. The dash-dotted curve shows the model Eq.\,\ref{eqn:energyEigenvalue} for $s=0.5$, inflecting at the trap-shifted resonance $\Deltat=\Delta_0$. All error bars represent \SI{68}{\percent} confidence intervals.} 
    \label{fig:sec2}
\end{figure*}

Successful imaging requires high photon scattering rates in addition to efficient cooling. Both processes are affected by the detunings of involved lasers from atomic resonances, more precisely the Raman two-photon detuning $\delta$ and the repumper detuning $\Delta$. Hence, good knowledge and control of the light shifts of the atomic levels during the imaging and cooling processes is essential. 

For continuous Raman sideband cooling (cRSC), the sideband transition frequency needs to resonantly connect motional states with different quantum numbers (see Fig.~\ref{fig:setup}). Therefore, we have to consider the light shifts induced by the repumper onto the ground state hyperfine levels. In the following, we use the notation $\ket{F, m_F}$ for a ground state ($\ket{F, m_F}'$ for an excited state) introduced in Sec.~\ref{sec:methods}. The upper hyperfine state $\ket{2,-2}$ is an uncoupled dark state to the $\sigma^-$ repumper, but the coupling to the lower hyperfine state $\ket{1,-1}$ results in a differential light shift $\deltals$ which modifies the resonance of the Raman transition. This shift grows when tuning the repumper close to resonance in order to increase the number of scattered photons for high imaging signal-to-noise ratios. 

The frequency of the cooling sideband $\deltac$ (Fig.~\ref{fig:setup}) is therefore shifted by an amount $\deltals$ with respect to the position of the carrier in the absence of light shifts.
Since the optimal detuning $\delta=\deltac$ (resonant addressing) depends on the parameters of the repumper, the optimization of imaging and cooling has to be carried out in a coupled 3D parameter space of Raman detuning $\delta$, repumper intensity $I_\text{rep}$, and repumper detuning $\Delta$. To tackle this problem, we characterize the repumper-induced differential light shifts prior to the imaging optimization.

As a model for the differential light shifts $\deltals$ we use a simple driven two-level system. The ground-state $\ket{1,-1}$ is coupled to the excited state $\ket{2,-2}'$ by the repumper light field, at Rabi frequency $\Omega$ and with a detuning $\Delta$. The system is described by the non-hermitian Hamiltonian, with energy eigenvalues $\mathcal{E}$ and light shifts $\deltals=\mathcal{E}/\hbar$ given to first order in the saturation parameter $s=2\left(\Omega/2\gamma\right)^2=I_\text{rep}/I_\text{sat}$ by
\begin{IEEEeqnarray}{rCl}
	\mathcal{\deltals} &\approx& \frac{\Delta}{2} \frac{\gamma^2}{\Delta^2 + \gamma^2} \, s + \mathcal{O}(s^2) \ . \label{eqn:energyEigenvalue}
\end{IEEEeqnarray}
In the low power limit ($s \ll 1$) or the large detuning limit ($\Delta \gg \gamma$) the induced light shift $\deltals$ scales linearly with intensity and follows a dispersive Lorentzian dependence on the detuning. Details on this calculation are given in Appendix~\ref{sec:ls_models}.

We calibrate the saturation parameter $s=I_\text{rep}/I_\text{sat}$ by monitoring saturation of the optical pumping rates between hyperfine ground states for a single atom as a function of the intensity $I_\text{rep}$. The repumper detuning with respect to free space also includes the AC-Stark shift induced by the dipole traps on the D1 transition. We define the total detuning with respect to the free-space resonance as $\Deltat=\Delta+\Delta_0$, with $\Delta$ the detuning from the trap-shifted resonance (used in the model) and $\Delta_0$ the dipole-trap induced light shift. For a trapped atom at the bottom of the lattice potential, the resonance condition is met for $\Deltat=\Delta_0$, with the shift $\Delta_0 /2\pi=\SI{32(4)}{\mega\hertz}$ calibrated independently with respect to a spectroscopy cell reference. 

We measure the differential light shift by detecting the displacement of the cooling sideband as a function of the saturation parameter Fig.~\ref{fig:sec2}(a) and the repumper detuning (b). We load a single atom into the resonator, precool it with degenerate Raman sideband cooling (dRSC)~\cite{Urunuela2020} to reach the Lamb-Dicke regime, pump it to the state $\ket{2, -2}$,  and then cool it with cRSC for a fixed time. We use the fiber quantum channel to probe the presence or absence of the atom inside the resonator, before and after the cooling slot. This measurement is repeated multiple times with different sets of parameters $\{\delta, s, \Deltat\}$ to map the survival probability to the parameter space for cRSC. 
For each set of repumper parameters $\{s, \Deltat\}$, the light-shifted sideband frequency $\deltac =\deltals + \nu$ is found as the value of $\delta$ maximizing the survival probability (inset of Fig.~\ref{fig:sec2}(a)), thus matching the Raman resonance condition $\delta=\deltac$ which provides the most efficient cooling. In Fig.~\ref{fig:sec2}(a) we show selected scans of the measured parameter map $\deltac(s, \Deltat)$, featuring the detuning $\deltac$ for three illustrative values of $\Deltat$, as a function of the repumper saturation parameter $s$, documenting the expected linear dependence on the repumper intensity, with the slope determined by the detuning. We observe that the repumper detuning at which the differential light shift becomes zero for all intensities occurs at $\Deltat/2\pi\approx \SI{28}{\mega\hertz}$, showing a small offset to the red with respect to the trap-shifted resonance calibrated independently. 

In Fig.~\ref{fig:sec2}(b) we have plotted the measurements of $\deltac$ as function of the repumper detuning $\Deltat = \Delta+\Delta_0$, for three selected values of $s$ showing dispersive curves with inflection points near resonance, at $\Delta/2\pi\approx \SI{-4}{\mega\hertz}$ and $\Deltat/2\pi\approx \SI{28}{\mega\hertz}$ (in agreement with the zero-shift line in Fig.~\ref{fig:sec2}(a)). For comparison we show the two-level model Eq.\,\ref{eqn:energyEigenvalue} for $s=0.5$. According to the model, the strongest differential light shifts ($\deltals=\deltac-\nu$) are expected for repumper detunings of $\Delta = \pm \gamma$, and the inflection point at resonance $\Delta=0$ (or $\Deltat=\Delta_0$).

The dispersive shape of the measured differential shift $\deltac$ is expected but deviates from the simple model of Eq.\,\ref{eqn:energyEigenvalue}. The deviation may be traced to several influences:
(1) The spatial distribution of atoms results in interactions with a range of laser beam intensities both in the optical traps, the Raman beams and the very narrow intra-cavity repumper beam which lead to a smoothing of the observed dispersive curve compared to the sharp model curve. (2) The technically challenging calibration of repumper intensity can underestimate the $s$-parameter leading to higher amplitudes than expected from Eq.\,\ref{eqn:energyEigenvalue}. (3) The data shows stronger shifts for blue repumper detunings ($\tilde{\Delta} > \Delta_0$) than for red ($\tilde{\Delta} < \Delta_0$), and the zero-shift condition at a smaller detuning than expected. We attribute these observations to detuning-dependent heating rates causing extended oscillations of the atoms in the trapping potential~\cite{Martinez-Dorantes2018}, analyzed further in Section~\ref{sec:imaging}.

\section{Balancing cooling and heating dynamics with imaging parameters}
\label{sec:imaging}

\begin{figure*}[t]
    \centering
    \includegraphics{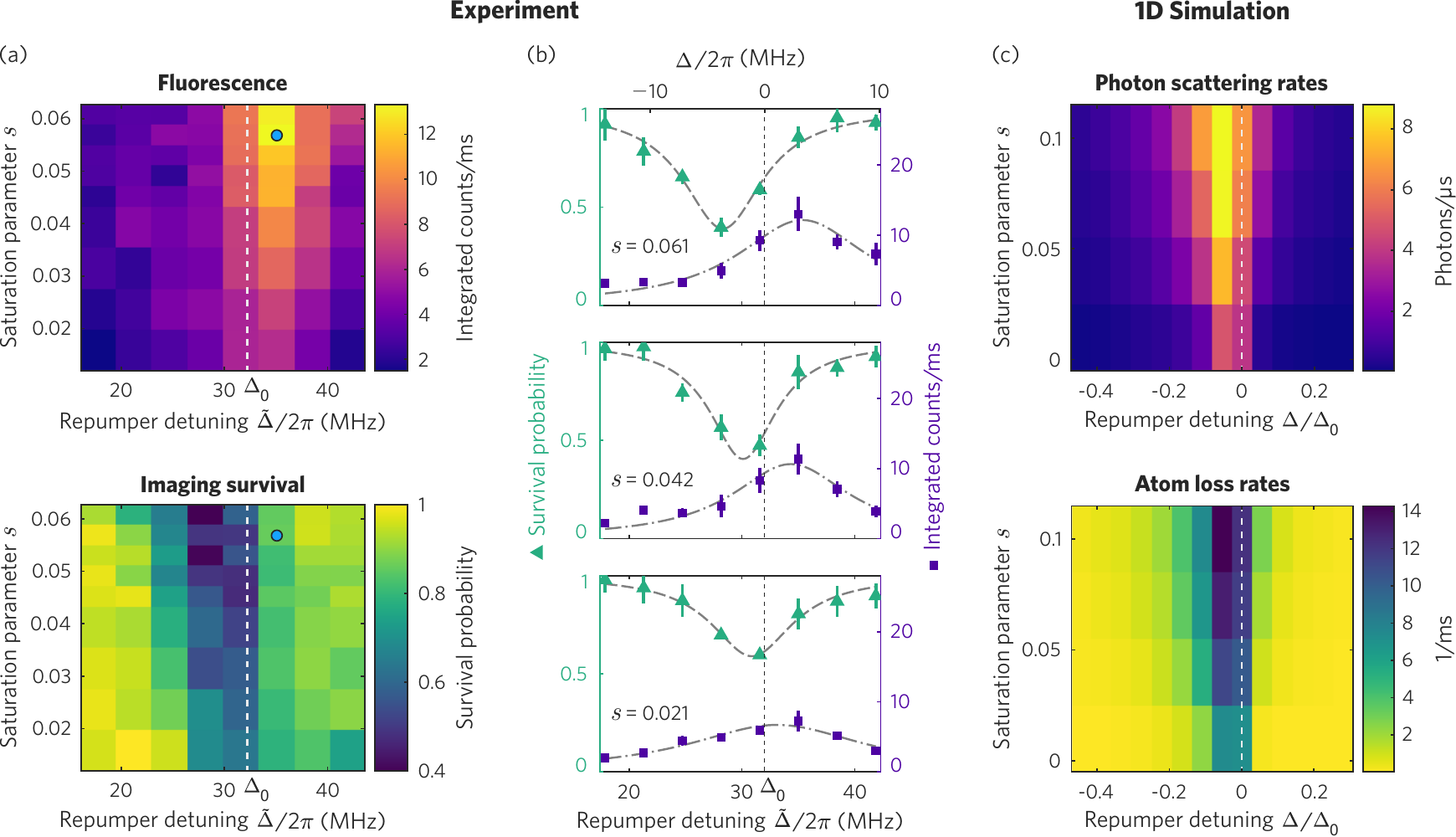}
    \caption{Balancing cooling-heating rates and fluorescence for imaging with cRSC. (a) Measured mean fluorescence per atom and survival probability for \SI{1}{\second} of cRSC, as function of repumper detuning $\Deltat$ and repumper saturation parameter $s=I_\text{rep}/I_\text{sat}$. The Raman addressing is kept resonant on the cooling sideband by compensating the differential light shifts induced by the repumper beam. The blue marker shows the optimum parameters at which the atom pictures in Fig.~\ref{fig:idea} were taken with $\text{SNR}\!\approx\!13$. (b) Cross sections of the fluorescence and survival maps versus detuning for selected intensities, shown with fitted Lorentz functions as a guide to the eye. The minimum in survival shifts towards red repumper detunings for an increase in intensity, contrary to the fluorescence peak that remains fixed close to resonance. We attribute the asymmetry of the survival behavior to DFFs-induced heating. (c) Photon scattering rates and atom loss rates computed with a 1D semi-classical Monte Carlo simulation of scattering in dressed-state potentials. The asymmetric shape of the atom loss rates validates DFFs as the source of heating in the measurement.}
    \label{fig:RamanImage}
\end{figure*}

The optimization of fluorescence imaging of atoms inside the resonator necessarily requires a trade-off of two competing processes in this imaging technique: (1) photon scattering that causes heating of the atom, and (2) Raman sideband cooling that counteracts the heating effect. As discussed in Sec.~\ref{sec:methods} (see also Fig.~\ref{fig:setup}(b)), the working principle of Raman sideband cooling relies on pumping the atom to the motional ground state (in full notation $\ket{2, -2 ;0}$), which is a dark state decoupled both from the Raman and the repumping light. On the other hand, continuous scattering of photons is required for imaging during the camera exposure time. Therefore, the atoms have to be heated out of the dark state but still kept close to the motional ground state for small atom loss. In \cite{Lester2014} parametric heating was used for this purpose, however in our case the intra-cavity lattice noise is sufficient to drive the atoms out of the ground state. 

For optimal imaging with continuous Raman sideband cooling, we find that the balance of heating-cooling rates and scattering can be tuned with the parameters of the repumper light. We have explored its effect on the heating-cooling trade-off and photon scattering rate, and optimize for a high SNR. The complexity of the parameter space is reduced by keeping the Raman two-photon cooling sideband on resonance using the differential light shift measurements during cRSC, analyzed in Sec.~\ref{sec:lightshifts}. Then, we simultaneously measure the imaging fluorescence intensity and the atom survival probability as a function of the repumper detuning and intensity. 

For the corresponding measurements we use an  experimental sequence similar to the light shifts measurements in Sec.~\ref{sec:lightshifts}: we begin with loading a single atom into the 3D lattice at the FFPC center, dRSC precooling, and initializing in the state $\ket{2, -2}$. Then, a first cavity-based atom detection is applied to probe the presence of an atom, followed by an EMCCD image acquisition under cRSC with \SI{1}{\second} exposure, at the end a second cavity detection is performed to verify the survival of the atom in the trap. 

The fluorescence signal is obtained by integrating the EMCCD counts over a $33\times\SI{33}{pixels}$ large image region corresponding to a $15\times\SI{15}{\micro\meter}$ large area at the center of the cavity. Reduced fluorescence due to atom loss is excluded by selecting the images based on successful presence and survival detections. Also, images with more than one atom are discarded by postprocessing with an atom detection algorithm. To quantify the balance of heating-cooling rates we take as a figure of merit the probability of a successful survival cavity-based detection conditioned on a prior successful presence detection before the imaging interval. 

The results for fluorescence intensity and imaging survival probability are shown as 2D color maps in  Fig.~\ref{fig:RamanImage}(a), where cross sections for selected $s$-values are given in panel (b). The combined information of fluorescence and survival maps (upper and lower panel, respectively) permits to identify a region favorable for imaging with $\text{SNR}\!\approx\!13$, of high fluorescence ($\sim\!\SI{2e4}{photons\per\second}$) and high survival ($>\!\SI{90}{\percent}$), at parameters $\Deltat/2\pi\!\approx\!\SI{35}{MHz}$ and $s\!\approx\!0.057$. These parameters were used to capture the images in Fig.~\ref{fig:idea}(b) with one, two and three atoms.

Qualitatively the behaviour of the fluorescence intensity (purple slices in \ref{fig:RamanImage}(b)) exhibits a symmetric lorentzian-like function as expected for a resonant spectrum with amplitude increasing with repumper intensity. We attribute a small $\sim\!\SI{3}{\mega\hertz}$ blue shift of the lorentzian center from the trap-shifted resonance to residual effects of atom loss for the cases of undetected more-than-one atom images.

The imaging survival map (Fig.~\ref{fig:RamanImage}(a) lower panel and green slices in Fig.~\ref{fig:RamanImage}(b)) shows an unexpected behavior. If considering only scattering-induced heating, we would expect the lowest survival rate on resonance, with the dip becoming deeper and broader for higher intensities. This is indeed the case at low intensities, i.e. for $s\!<\!0.02$. However, for higher intensities the survival dip experiences a clear shift towards red detunings with a linear dependence on the intensity. This results in an asymmetric V-shape of the imaging survival map with higher survival probabilities on the blue detuning side than on the red. The model of the atom motion inside the trap for finite temperature introduced in Sec.~\ref{sec:lightshifts} predicts a constant red shift, but not a power-dependent shift. The observed asymmetric behavior indicates heating rates induced by the repumper beams that depend both on their detuning and intensity. 

During the scattering cycles induced by the repumper field, the atoms undergo rapid transitions between ground state and and excited state, which are associated with trapping and anti trapping potentials, respectively. For our \SI{868}{\nano\meter} lattice the estimated polarizability ratio of the $^{87}$Rb excited state $5^2 \text{P}_{\nicefrac{1}{2}}$ over the ground state $5^2 \text{S}_{\nicefrac{1}{2}}$ is $\chi\approx-0.59$ \cite{Arora2012}. We therefore attribute this asymmetry to dipole-force fluctuations (DFFs) \cite{Taieb1994,Martinez-Dorantes2018}.

To gain insight into the effect of DFFs in our measurements, we set up a one dimensional semi-classical Monte Carlo simulation of the scattering process of an atom trapped in a lattice potential under near-resonant illumination, based on \cite{Martinez-Dorantes2018}. 
The simulation combines the classical motion of an atom in the dressed-state potentials, with the position-dependent transition rates between dressed states and the corresponding probabilities of scattering events. For an ensemble of atoms with a set of repumper parameters, it calculates the time-dependent photon emission rate and the increase of mean kinetic energy. From the time evolution of these values we extract the mean scattering rate and the exponential loss rate of atoms escaping from the trap.
More details on the theory of DFFs and on the Monte Carlo simulation are presented in Appendix~\ref{sec:DFFs}.

In Fig.~\ref{fig:RamanImage}(c) we show the simulated 2D maps of atom loss rates and photon scattering rates as a function of repumper detuning and intensity. A qualitative comparison with the results of the measurement in Fig.~\ref{fig:RamanImage}(a) shows the striking similarities between the simulated loss rates and the measured survival probabilities. Moreover, the map of photon scattering is in good agreement with the detected fluorescence. 
Our model supports DFFs as the main effect governing the heating dynamics that lead to the observed asymmetric survival rates. For a compromise of high fluorescence and at the same time high survival probability with the repumper close to resonance, a red detuning should be avoided, while a blue detuning can prevent DFFs-induced heating. This holds for the cases where the AC-Stark shift induced by dipole traps increases the energy of the atomic transition.
Such parameter-dependent heating rates also offer an explanation for the asymmetric deformation of the light-shifts measurements previously shown in Fig.~\ref{fig:sec2}(b).

\section{Conclusion}
\label{sec:conclusion}

In summary, we have shown that imaging of single atoms trapped inside a high-bandwidth FFPC can be successfully implemented, despite the strong Purcell effect. Note that the fluorescence suppression by the Purcell effect is closely related to the observation of a glowing laser medium being visually dimmed when the laser mode ignites. Our experimental realization takes advantage of our continuous Raman cooling scheme by using the scattered repumper photons, which are not resonant with the cavity. Our imaging technique requires only a single free-space beam together with intra-cavity fields, making it ideal for situations with limited optical access, e.g. in miniaturization trends in quantum technologies.

The signal-to-noise ratio in our experiment is limited by background scattering from the cavity mirrors. In contrast to other experiments relying on large detunings \cite{Cheuk2015,Martinez-Dorantes2018}, this enforces low power intensities and small detunings to obtain a high SNR. Here, we have demonstrated an imaging SNR of 13 based on a compromise between photon scattering rate and three-dimensional sideband cooling efficiency. During the imaging process more than \SI{85}{\percent} of the population remains in the motional ground state per direction, sufficient for high fidelity detection of single atoms in the optical lattice. The optimization of the relevant parameters was enabled by a detailed spectroscopic investigation during cRSC. This also shed light onto deviations from the expected light shifts, revealing a parameter-dependent heating mechanism that we attributed to dipole-force fluctuations impairing the atom survival at red detunings, but avoided at blue detunings.
This was further validated by a Monte Carlo simulation of the atom's scattering dynamics in the position-dependent dressed-state potentials. The observed DFFs heating effect is not limited to our particular FFPC setup with cRSC. On the contrary, it is expected to be a relevant effect in most experiments involving near-resonant illumination of atoms trapped in a lattice, depending on the polarizabilities of the states involved.

While global positioning of the atomic ensemble is controlled with our optical conveyor belt, the imaging technique shown in this work represents an important step towards ultimate position control and manipulation of single atoms in miniature optical cavities. It enables to determine the number and position of the atoms within the resonator in a non-destructive manner, and paves the way for creating atomic arrays with predefined number and positions in the cavity. This can be implemented in the experiment by integrating single-atom addressing optical tweezers controlled by a spatial light modulator.

\begin{acknowledgments}
This work has been funded by the Deutsche Forschungsgemeinschaft (DFG, German Research Foundation) with Project No. 277625399-TRR 185 \textit{OSCAR}, under Germany's Excellence Strategy – Cluster of Excellence Matter and Light for Quantum Computing (ML4Q) EXC 2004/1 – 390534769, as well as by the Bundesministerium f\"ur Bildung und Forschung (BMBF), project FaResQ. We thank T. Macha, D. Pandey and M. Martinez-Dorantes for insightful discussions and technical support in the early stage of the presented work.
\end{acknowledgments}


\appendix

\section{LINEWIDTH-REDUCTION PREPARES DBR LASER FOR PHASE LOCKING}
\label{sec:Raman-LRDBR}

 \begin{figure*}
    \centering
    \includegraphics{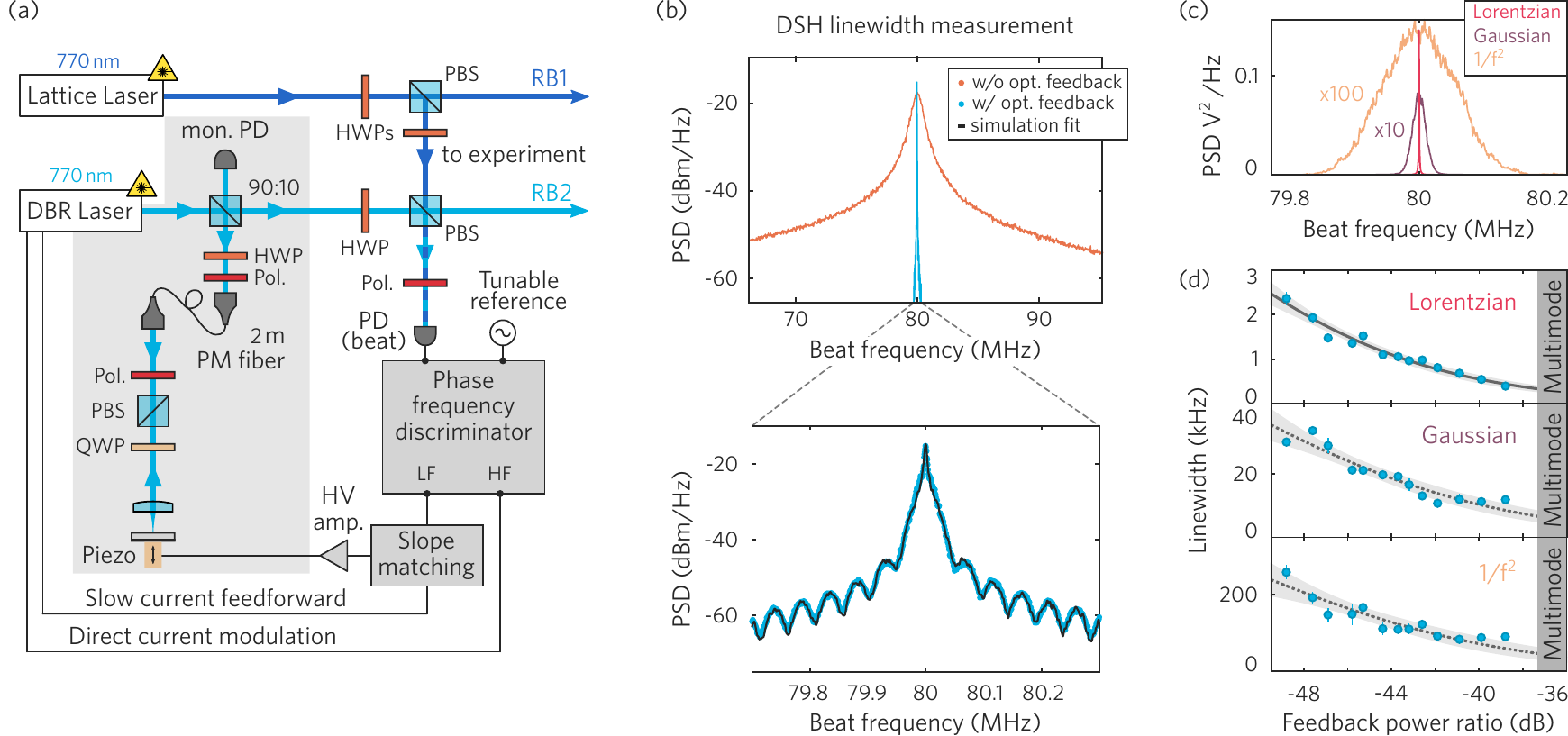}
     \caption{(a) Experimental setup for linewidth reduction of the DBR laser RB2 and phase locking to the lattice laser RB1. (b) Delayed self heterodyne (DSH) spectrum of the DBR laser with and without optical feedback (feedback power ratio of \SI{-40}{dB}). A linewidth reduction in two orders of magnitude can be observed when the laser is subject to optical feedback (blue curve). Lower panel: close-up of the DSH spectrum and simulation based fitting to extract linewidth components. (c) Pure line-shapes corresponding to the three laser frequency-noise components, namely  white (Lorentzian), flicker (Gaussian) and random-walk frequency noise ($1/f^2$) extracted with the simulation-based fitting routine. (d) Estimated  linewidth corresponding to each frequency-noise component as a function of the feedback power ratio.}
    \label{fig:linewidthReduc}
\end{figure*}

The two-photon Raman cooling transitions are driven in our setup by the \SI{770}{\nano\meter} lasers RB1 and RB2 in Fig.~\ref{fig:setup}(b) with frequency difference $\sim\!\SI{6.8}{\giga\hertz}$. The primary RB1 laser field is drawn from an interference filter stabilized diode laser \cite{Baillard2006} with rms-linewidth of order few kHz. It is injected into the FFPC and serves not only as a blue detuned optical lattice for atom trapping but also to simultaneously stabilize the cavity length at the atomic resonance line at \SI{780}{\nano\meter}. For this purpose its wavelength is referenced to an optical frequency comb.

In order to obtain sufficiently high resolution for driving coherent Raman transitions between hyperfine levels the secondary \SI{770}{\nano\meter} Raman beam (RB2) is phase locked to RB1. We have chosen a distributed Bragg reflector (DBR) laser which offers convenient and mode hop free tuning over several GHz, and good long term frequency stability. The drawback of DBR components, however, is a generally large Schawlow-Townes-Henry linewidth \cite{Henry1982} of order a few hundred \si{\kilo\hertz}, too broad for realizing sufficient feedback bandwidth for phase locking. It is known that external optical feedback can help to reduce the DBR linewidth  \cite{Agarwal1984}, a road we have followed here. 

We outline the application of optical feedback from a \SI{2}{\meter} long external feedback path to reduce the linewidth of a DBR laser diode based on the work in \cite{Lin2012}. We use a delayed self-heterodyne (DSH) linewidth measurement and a simulation-based fitting routine \cite{Ma2019} to analyze and estimate all the linewidth components under the influence of varying optical feedback strengths. A current feed-forward was implemented to achieve mode-hop-free tuning of the laser frequency over the free spectral ranges of the external cavity. With that, continuous tuning of up to \SI{2}{\giga\hertz} was achieved. Furthermore, we present the phase locking setup for Raman sideband cooling in our experiment in Fig.~\ref{fig:linewidthReduc}(a). 

To reduce the Lorentzian linewidth of the \SI{770}{\nano\meter} DBR laser, an external optical feedback path is implemented (highlighted in grey in Fig.~\ref{fig:linewidthReduc}(a)). Approximately \SI{4}{\percent} of the emitted laser light reflected at the 90:10 beam splitter (BS) is coupled into a \SI{2}{\meter} long polarization maintaining (PM) fiber. From the fiber output, the laser light passes a quarter wave plate (QWP), a polarizing beam splitter (PBS) and a polarizer (Pol.), and is focused on a mirror mounted on a piezo using a lens in cat’s eye configuration. The light reflected at the mirror is then coupled back into the fiber and fed into the laser diode. The polarization optical elements are used to adjust the feedback power. The monitor photodiode (mon. PD) placed at the free arm of the 90:10 BS is used to monitor the feedback power. At a fixed external feedback path length of \SI{2}{\meter}, the feedback power ratio (ratio of the power entering back into the DBR laser to emitted power) can be varied from \SI{-50}{dB} to \SI{-30}{dB}. 

To measure the linewidth of the laser subject to varying optical feedback levels, we use a DSH method first described in~\cite{Okoshi1980}. It consists of a Mach-Zehnder type interferometer, where the light passing one of the arms is frequency shifted by $\delta_{f}=\SI{80}{\mega\hertz}$ and the other passing a \SI{4.9}{\kilo\meter} optical fiber delay. The spectrum of the beat signal of the recombined arms is analyzed. Fig.~\ref{fig:linewidthReduc}(b) shows a DSH spectrum with and without the optical feedback. When the laser is not subject to optical feedback (orange curve in top panel of Fig.~\ref{fig:linewidthReduc}(b)), the DSH spectrum is a broad gaussian curve with an estimated linewidth of $\sim\!\SI{700}{\kilo\hertz}$. With optical feedback, the width of the DSH spectrum reduces by at least two orders of magnitude (blue curve, top panel of Fig.~\ref{fig:linewidthReduc}(b) and close-up view in lower panel). The spectrum shows a Lorentzian at the modulation frequency with periodic ripples on the wings indicating the residual coherence between the interferometer arms.

To extract the different linewidth components from the spectrum, we use a simulation-based fitting routine based on \cite{Ma2019}. It emulates the experimentally recorded spectra by simulating the photodiode signal with numerically generated phase noise contributions. The phase noise signal $\phi(t)$ is generated by the superposition of a white $\phi_{w}(t)$, flicker $\phi_{f}(t)$ and random-walk frequency noise $\phi_{r}(t)$ source with $\mathcal{A}_{w}$, $\mathcal{A}_{f}$ and $\mathcal{A}_{r}$ as the amplitudes of the corresponding noise sources: 
\begin{eqnarray}
\phi(t) = \mathcal{A}_{w}\phi_{w}(t)+\mathcal{A}_{f}\phi_{f}(t)+\mathcal{A}_{r}\phi_{r}(t)\,.
\label{eqn:PhaseNoiseSeq}
\end{eqnarray}
The power spectral density (PSD) of the heterodyne signal is calculated and the noise amplitudes are adjusted to fit the numerically simulated spectrum to the experimental data using a Nelder-Mead simplex optimization. Fig.~\ref{fig:linewidthReduc}(c) shows exemplary simulated spectra of the different noise components. Our code for the simulation-based linewidth measurement is openly available in the repository \cite{SimBasedLinewidth}. 

Once the noise parameters are found, the Lorentzian linewidth $\delta\nu$ of the laser is estimated by fitting the PSD of white phase noise by $\mathcal{S}_{w}(f)=\frac{\delta\nu}{\pi f^2}$. The fitted Lorentzian linewidth of the laser as a function of the feedback power ratio is shown in the top panel of Fig.~\ref{fig:linewidthReduc}(d). Lorentzian linewidths as low as \SI{500}{\hertz} are reached with optical feedback ratio close to \SI{-36}{dB}. With larger feedback powers the laser becomes multi-mode as also observed in ~\cite{Lin2012}. The simulation based fitting also allows to extract weak components of higher order frequency noise. Here, we include flicker noise (1/$f$), which results in a Gaussian lineshape, and frequency random-walk noise (1/$f^2$). The retrieved linewidth for both with respect to feedback power is presented in the lower two panels of Fig.~\ref{fig:linewidthReduc}(d).

For the experiment we fixed the feedback ratio at \SI{-43}{dB}, where the Lorentzian linewidth is sufficiently low for our purposes. The flicker and the random-walk noise are suppressed using a fast phase lock, with the setup shown in Fig.~\ref{fig:linewidthReduc}(a). The two lasers are combined on a polarizing beam splitter (PBS) and the corresponding beat signal is detected on a photodiode (PD beat). A phase frequency discriminator (PFD) is used to compare the beat signal of the linewidth-reduced DBR laser and RB1 to a tunable stable reference to quantify frequency and phase deviations. The low frequency (LF) and the high frequency (HF) error signals are further processed to phase lock the two lasers. The HF error signal is used to compensate for the fast phase fluctuations. Its corresponding feedback is directly applied to the laser diode current via a loop filter. The LF error signal is used to correct for slow frequency drifts ($\sim\!\si{\kilo\hertz}$). Its corresponding feedback signal is applied to the piezo of the mirror at the end of the external feedback path after being amplified by a high voltage amplifier. As the \SI{2}{\meter} long external feedback path has a corresponding free spectral range of \SI{56}{\mega\hertz}, the frequency tunability is regained by a current feed-forward. Currently, the mode-hop free tuning range of up to \SI{2}{\giga\hertz} is limited by the maximum stroke of the piezo of the external feedback mirror.

\section{3D CONTINUOUS RAMAN SIDEBAND COOLING WITH ONE FREE-SPACE BEAM ONLY AND INTRA-CAVITY FIELDS}
\label{sec:rsc3d}

\begin{figure}
    \centering
    \includegraphics{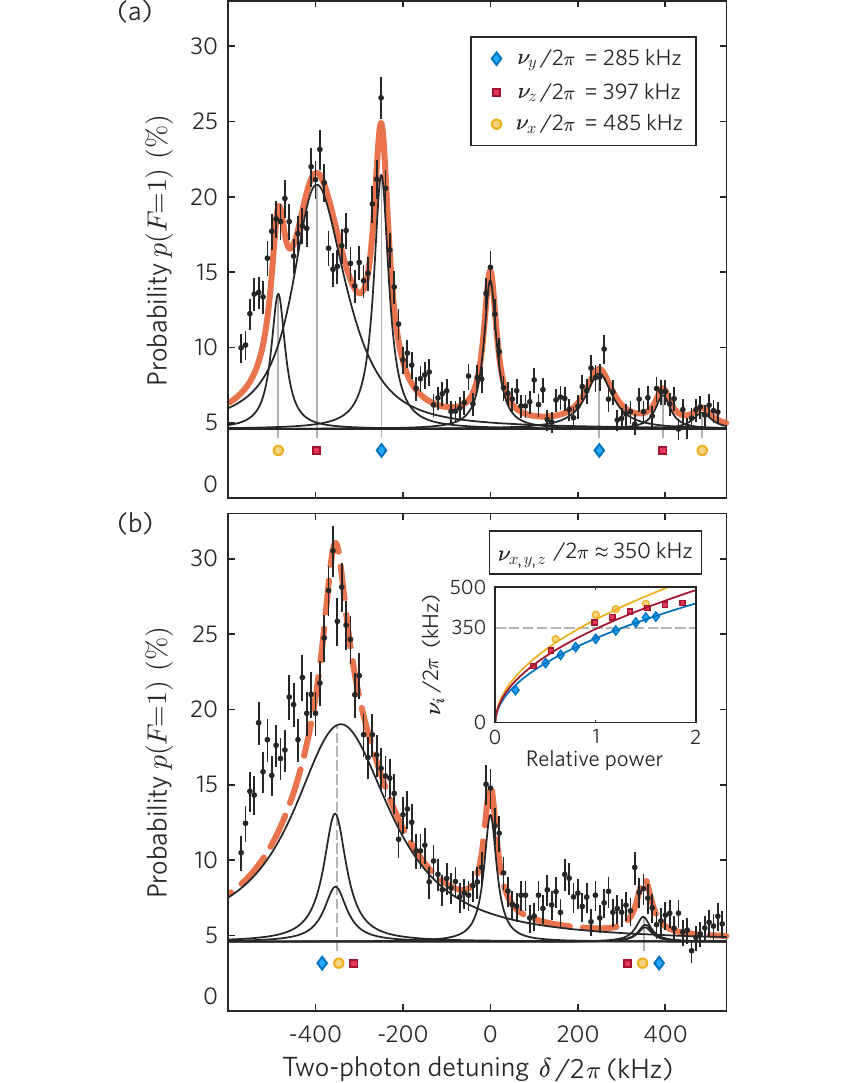}
    \caption{Raman spectra of an atom trapped in the 3D lattice inside the FFPC. (a) For unequal trapping frequencies the first order heating ($\delta<0$) and cooling ($\delta>0$) sidebands corresponding to each lattice dimension can be identified. The trap frequencies $\nu_i$ are extracted by fitting the data with the sum of seven Lorentzians (orange line, individual peaks in black). The carrier transition is highly suppressed as a consequence of using the blue-detuned intra-cavity lattice as a Raman beam \cite{Reimann2014,Neuzner2018}. (b) Raman spectrum after independently tuning the lattice beam intensities (inset) superposes all three trap frequencies (sidebands) into near-degeneracy around \SI{350}{\kilo\hertz} (the dashed orange curve serves as a guide to the eye). The asymmetry of the sidebands proves the efficiency of the 3D-cRSC scheme ($>\!\SI{85}{\percent}$ ground state fraction).} \label{fig:raman_spectra}
\end{figure}

In Section \ref{sec:methods} we summarized our continuous Raman sideband cooling (cRSC) scheme focusing on the 1D case. Here we explain in detail how we extend this scheme to cool with respect to all directions of our 3D lattice, with only a single free-space beam (RB2) and the remaining beams as intra-cavity fields (RB1 and repumpers). For details on the setup see Fig.~\ref{fig:setup}. To drive the cooling transitions corresponding to all lattice dimensions simultaneously, the key point is to bring the relevant motional oscillator frequencies close to degeneracy in all three directions.

We evaluate Raman spectra of single atoms trapped in our cavity in order to identify the different heating and cooling sidebands. A single atom is loaded into the cavity, precooled with degenerate Raman sideband cooling (dRSC)~\cite{Urunuela2020}, and prepared into $\ket{2, -2}$. Next, a \SI{500}{\micro\second} Raman pulse with variable two-photon detuning $\delta$ drives Rabi oscillations between the states $\ket{2, -2}$ and $\ket{1,-1}$, after which the state of the atom is probed by means of non-destructive cavity probing \cite{Gallego2018}. Within the trapping lifetime of about one minute, the same atom can be reinitialized, cooled, and recycled by applying \SI{5}{\milli\second} of cRSC at $\delta/2\pi=\SI{350}{\kilo\hertz}$ before the spectroscopy pulse, for up to 600 measurement cycles with different $\delta$ values.  Additionally, we use the cavity detection to postselect the measurements upon the presence of an atom before and after each spectroscopy sequence.

We record the Raman spectra  by monitoring the probability of an atom to be transferred to $\ket{F=1}$ as a function of the Raman two-photon detuning. In Fig.~\ref{fig:raman_spectra}(a) we show the spectrum recorded before overlapping the sidebands. The error bars of the data points represent the $\SI{68}{\percent}$ confidence interval of the binomial distribution of each measurement. A fit to the spectrum using seven Lorentzian curves allows us to extract the oscillation frequencies corresponding to each lattice dimension. After intensity adjustment of the lattice beams the spectrum Fig.~\ref{fig:raman_spectra}(b) shows almost degenerate sideband frequencies.

On the left side of the spectra ($\delta<0$) we identify the heating sidebands ($\Delta n = +1$) for each lattice dimension and the corresponding cooling sidebands ($\Delta n = -1$) on the right side ($\delta>0$). Note that for Fig.~\ref{fig:raman_spectra}(a) cooling the atoms with dRSC and off-resonant cRSC was sufficient to allow the measurement. As a consequence of using a blue detuned intra-cavity lattice as Raman beam RB1 the atoms are trapped near the anti-nodes which causes the carrier transition ($\Delta n = 0$) at $\delta=0$ to be highly suppressed \cite{Reimann2014,Neuzner2018}. The broad shape of the $z$ heating sideband originates from inhomogeneous broadening due to the radial position distribution of the atoms in the narrow-waist intra-cavity lattice.
The $z$-distribution is negligible since the trapping occurs within the Rayleigh length of the cavity mode. The non-zero spectrum baseline reflects a $\ket{2,-2}$ state preparation efficiency of $\sim\!\SI{95}{\percent}$.

Based on these measurements we calibrate the displacement of each sideband for different intensities of the lattice beams (inset of Fig.~\ref{fig:raman_spectra}(b)). We typically tune all three trap frequencies to $\nu_{x,y,z}/2\pi \approx \SI{350}{\kilo\hertz}$ shown in the spectrum of Fig.~\ref{fig:raman_spectra}(b). Setting the Raman beams detuning to $\delta/2\pi=\SI{350}{\kilo\hertz}$ and switching the repumper beams (Fig.~\ref{fig:setup}) on simultaneously, cRSC is active along all lattice dimensions, as discussed in Section \ref{sec:methods}. From the ratio of heating-cooling sideband amplitudes we estimate a ground state occupation of more than \SI{85}{\percent} in all lattice dimensions, corresponding to mean motional excitations $\overline{n}\approx 0.17$ and a residual temperature of \SI{1.4}{\micro\kelvin}.

\section{JUSTIFICATION AND LIMITATIONS OF THE TWO-LEVEL LIGHT SHIFT MODEL}
\label{sec:ls_models}

\begin{figure*}
    \centering
    \includegraphics{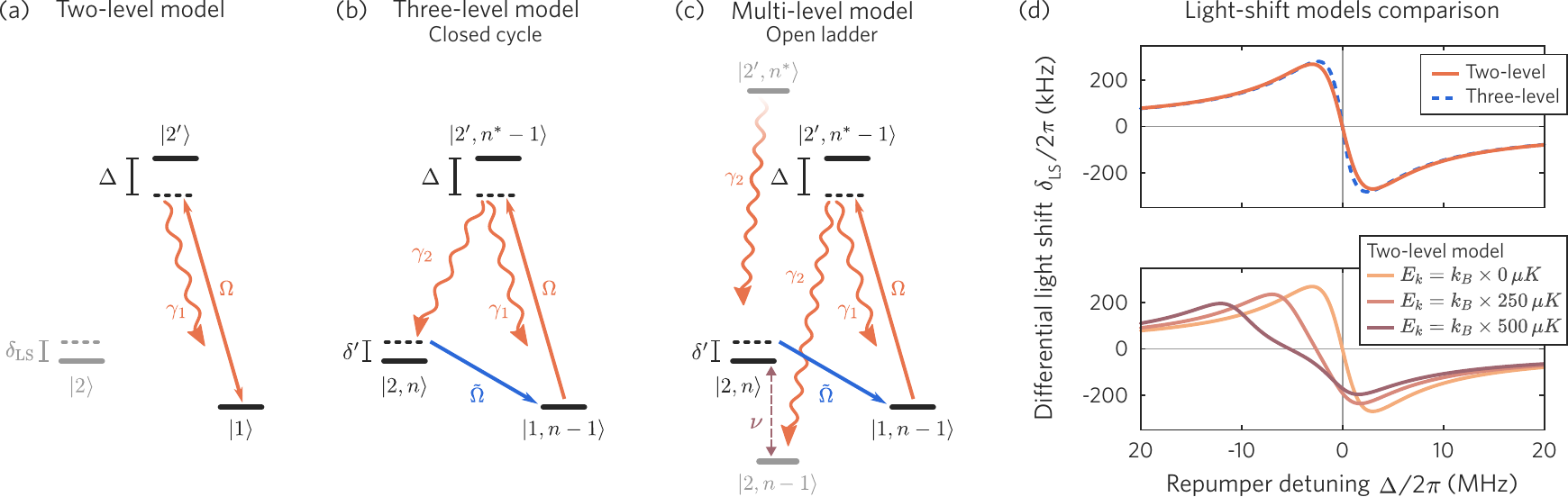}
    \caption{Differential light shifts models. (a) Two-level model used in the main text to fit the experimental data. (b) Three-level model including the Raman Rabi coupling \cite{hilker2017} for closed cycle treatment (c) Adjacent motional and relevant atomic levels (d) Comparison of the two light-shifts models (upper panel) for $^{87}$Rb, and of the two-level model when considering atomic oscillations in the lattice for different kinetic energies of the trapped atoms (lower panel). All of the plots consider a repumper saturation parameter $s=0.36$ and a trap depth $U_0=k_B\!\times\!\SI{0.5}{\milli\kelvin}$.}
    \label{fig:model_comparison}
\end{figure*}

In Section \ref{sec:lightshifts} of the main text we discuss the measurement of the differential light shifts (Fig.~\ref{fig:sec2}) as a function of repumper intensity and detuning during cRSC. We analyze our data with respect to a simple two-level model in the low power limit.

While the linear dependence on the power in Eq.\,\ref{eqn:energyEigenvalue} agrees with the experiment (Fig.~\ref{fig:sec2}(a)), the dispersive relation of repumper detuning $\Delta$ vs. $\deltac$ shows clear deviations from the measurement (Fig.~\ref{fig:sec2}(b)). Although a rough dispersive shape is observed, the measured relation is broadened and shows an asymmetric distortion towards red detunings of the repumper. The deviations from the  two-level model can be caused by a multitude of both experimental factors and physical effects. Here we detail the validity of the used model and some possible effects affecting the observed light shifts.

The two-level model focuses on the repumper coupling only and ignores the Raman interaction. We also assume that all atoms are subject to the same laser intensities. The ground-state $\ket{1,-1}=\ket{1}$ (to be short) is coupled to the excited state $\ket{2',-2}=\ket{2'}$ with repumper Rabi frequency $\Omega$ and detuning $\Delta$, while the second ground state $\ket{2,-2}=\ket{2}$ is considered a dark state for $\sigma^-$ polarization (grey levels in  Fig.~\ref{fig:model_comparison}(a)). The system is described by a non-Hermitian Hamiltonian~\cite{moiseyev1998,gardiner2004}, which in matrix representation reads as
\begin{IEEEeqnarray}{rCl}
	H &=& \hbar \, \begin{pmatrix}
		-\Delta - i\gamma & \Omega/2 \\
		\Omega/2 & 0
	\end{pmatrix} \label{eqn:twoLevelHamiltonian}
\end{IEEEeqnarray}
where the complex detuning $i\gamma$ introduces the finite linewidth of the optical transition. The light shift $\deltals$ for this two-level system is obtained by computing the energy eigenvalues $\mathcal{E}=\hbar\cdot\deltals$
\begin{IEEEeqnarray}{rCl}
	\mathcal{E} &=& \frac{\hbar}{2} \; \Re\left(-\Delta-i\gamma + \sqrt{(2s-1)\gamma^2 + \Delta^2 + 2i\gamma\Delta}\,\right), \qquad \label{eqn:eigenenergiesTwoLevel}
\end{IEEEeqnarray}
where $\Re$ denotes the real part, $2\gamma \approx 2\pi \cdot \SI{6}{\mega\hertz}$ the linewidth of the transition, and $s=2\left(\frac{\Omega}{2\gamma}\right)^2=\frac{I_\text{rep}}{I_{\text{sat}}}$ the saturation parameter. To first order in $s$ ($\Omega \ll 2\gamma$) or large detuning ($\Delta \gg 2\gamma$), the light shift becomes
\begin{IEEEeqnarray}{rCl}
\deltals &=& \frac{\mathcal{E}}{\hbar} \approx \frac{1}{2} \gamma^2 \frac{\Delta}{\Delta^2 + \gamma^2} \, s + \mathcal{O}(s^2) \label{eqn:eigenenergiesTwoLevelApprox}
\end{IEEEeqnarray}
which is Eq.\,\ref{eqn:energyEigenvalue} in the main text in Sec.~\ref{sec:lightshifts}.

In order to investigate the validity and limitations of the two-level approximation we explore the role of the Raman coupled second hyperfine level for our model (Raman Rabi frequency $\Omegat$, Fig.~\ref{fig:model_comparison}(b)). This scheme was first presented in \cite{hilker2017} and we will show that it modifies the calculated light shifts only marginally. In Fig.~\ref{fig:model_comparison}(c)) we include, for the sake of completeness, adjacent relevant motional levels for our discussion of the Raman cooling process, yielding the characteristic open ladder of lowering motional states. 

The motional atomic energy levels relevant for the Raman cooling process have quantum numbers $n$ and energy spacing $\hbar\nu$, see Figs.~\ref{fig:model_comparison}(b) and (c). A cooling cycle starts in the state $\ket{2 ,n}$, i.e. with an atom with motional excitation $n$. Raman coupling transfers the atom with Rabi frequency $\Omegat$ and  two-photon detuning $\delta'$ into the state $\ket{1 ,n-1}$ and thereby reduces the number of motional excitations by one. The Raman two-photon detuning is denoted with $\delta' =\delta -\nu$ here, since we consider only the Raman cooling sideband and neglect carrier and heating transitions. Optical repumping occurs at a Rabi frequency $\Omega$ and with detuning~$\Delta$, via the excited state $\ket{2' ,n^*-1}$. Note that the excited state (excitation number $n^*$) experiences a different trapping potential than the ground-states. In the Lamb-Dicke regime, where scattering events changing the motional state $n$ are unlikely, the excited state decays either back into the state $\ket{1 ,n-1}$ at a rate $\gamma_1$, or into the state $\ket{2 ,n-1}$ at a rate $\gamma_2$. In the latter case the cooling cycle is completed and the atom was cooled from the $n$-manifold into the $(n-1)$-manifold. Following \cite{hilker2017}, we note that the cooling occurs through the scattering mediated via the excited state $\ket{2' ,n^*-1}$, and thus the cooling rate is proportional to the population of the excited state $\rho_{ee}$, as long as the system has not yet reached the dark motional ground state.

To quantify the continuous Raman cooling rate as a function of the repumper parameters, the approach in \cite{hilker2017} is to find a steady-state solution $\rhoss$ for the excited state population. Under the assumptions of negligible population of the motional ground state and of balanced decay rates into and out of each $n$-manifold, the additional adjacent levels of the open ladder system Fig.~\ref{fig:model_comparison}(c) can be neglected, rendering the closed-cycle three-level model in Fig.~\ref{fig:model_comparison}(b) the next approximation  beyond the two-level model in Fig.~\ref{fig:model_comparison}(a).

Using a Lindblad master equation approach and for $\Gamma \gg \Omega \gg \Omegat$, with the atomic linewidth $\Gamma = 2 \left(\gamma_1 + \gamma_2\right)$, the steady state excited state population is approximated as
\begin{IEEEeqnarray}{rCl}
	\rhoss \approx  \label{eqn:rho33steadyStateApprox} & 
	& \frac{\Omega^2\,\Omegat^2}{2\Omegat^2\left(\Gamma^2 + 4 \Delta^2\right) + 4 \alpha \delta'^2 \Gamma^2 + \alpha \left(\Omega^2 + 4\delta'\tilde{\delta}\right)^2}  \qquad 
\end{IEEEeqnarray}
where $\tilde{\delta} = \Delta-\delta'$ is the relative detuning and $\alpha = \frac{\gamma_2}{\gamma_1 + \gamma_2}$ the effective repumping efficiency. Details of the derivation can be found in \cite{hilker2017}. Since the survival probability of atoms in the experiment is proportional to the cooling rate and in turn proportional to $\rhoss$, we look for the two-photon detuning $\deltam$ that maximizes the excited state population, therefore matching the differential light shift $\deltals = \deltam$. By imposing the condition $\frac{\partial \rhoss}{\partial \delta'} = 0$, we find that $\deltam$ obeys 
\begin{IEEEeqnarray}{rCl}
	0 =& \\ \label{eqn:expressionBestDelta}
	& \left(\deltam\right)^3 - \frac{3}{2} \Delta \left(\deltam\right)^2 + \left[\frac{\Delta^2}{2} - \frac{\Omega^2}{4} + \frac{\Gamma^2}{8}\right] \deltam + \frac{\Delta \Omega^2}{8} . \nonumber
\end{IEEEeqnarray}
The solution depends on the repumper Rabi frequency $\Omega$ and the repumper detuning $\Delta$, and using the relation $\deltam = \deltals = \deltac -\nu$ it can be compared to the two-level model (Eq.~\ref{eqn:eigenenergiesTwoLevelApprox}) used in the main text, and to the  measured quantity $\deltac$ (see Sec.~\ref{sec:lightshifts}). 

The comparison of the differential light shifts $\deltals$ calculated with both models is shown in the upper panel of Fig.~\ref{fig:model_comparison}(d)
for equal parameters. The results deviate slightly in the central region between the extrema, with the three-level model showing only a slightly sharper dispersive shape in the limit of low repumper intensities. Therefore, explicit accounting for the Raman cooling cycle does not explain the discrepancy to the experimentally determined light shifts. Even though the more elaborate model from \cite{hilker2017} gives insight into the Raman cooling rates in the full parameter region (detuning and intensity from the repumper and the Raman beams), we find that the two-level model suffices to describe the light shifts in the low intensity regime of our experiment with respect to light matter interactions.

As a consequence, the observed broadening and asymmetric distortions of light shifts towards red detuning have to be attributed to effects that are not included in the theoretical description of the Raman cooling process including: frequency fluctuations of the laser sources, spatial inhomogeneities of laser intensities and both statistical and thermal distribution of atoms among and within their lattice sites.

Laser frequency fluctuations of the repumper beam lead to a broadening, or flattening of the observed dispersive curve. The magnitude of this noise in our experiment is on the order of few $\SI{}{\mega\hertz}$ with a Gaussian distribution. The broadening would be symmetric since it corresponds to a convolution of the idealized measurement with the Gaussian but does not explain the asymmetric distortion towards red detunings. 

The repumper is conveniently coupled through one of the fiber cavity mirrors which, however, has the drawback to create a narrow beam profile, leading to a highly inhomogeneous repumper intensity distribution sensed by the atoms. The result is an effective reduction of the average intensity in the measurements which leads to a flattening of the measured curve compared to the model for a given repumper intensity, but again would not introduce any asymmetric distortion. 

The above described two and three-level models assume that an atom is fixed at he bottom of the trapping potential, and therefore experiences a fixed light shift of the optical repumping transition. Atoms with finite kinetic energy will, however, perform oscillations in the lattice site and therefore be exposed to varying light shifts within an oscillation period \cite{Dalibard1985,Taieb1994,Martinez-Dorantes2018}. In the harmonic approximation, the time-dependent position of an atom inside a lattice well is given as 
\begin{equation}
    x(t) = \left(\frac{2 E_k}{m \nu^2}\right)^{1/2} \, \sin \left(\nu \, t\right)
\end{equation}
with, $m$ the atomic mass, $E_k$ the kinetic energy and $\nu$ the angular trapping frequency. The corresponding light shift of the optical repumping transition leads to a position-dependent detuning
\begin{equation}
    \Delta(x) = \Delta + \frac{1}{2 \hbar} m \, \nu^2 \, x^2 \,\left(1-\chi\right)
\end{equation} 
where $\Delta$ is the AC-Stark-shifted detuning, meeting the resonance condition $\Delta=0$ at the bottom of the trapping potential, and $\chi$ is the ratio of polarizabilities of ground and excited state ($\chi=\alpha_e/\alpha_g$). In particular, for small red detunings there are positions in the atom oscillation trajectory where the resonance condition is reached ($\Delta(x) = 0$), which is not possible for blue detunings. We note that this holds for all cases where the AC-Stark shift increases the energy of the atomic transition. This process can be taken into account by averaging the differential light shift over one oscillation period
\begin{IEEEeqnarray}{rCl}
	\bar{\delta} &=& \frac{\nu}{2 \pi} \, \int_{0}^{\frac{2\pi}{\nu}} \deltals(x(t)) \,\text{d}t \label{eqn:lightShiftOscillationInTrap} \\
	&=& - \frac{1}{2} \, \gamma^2 \, \Re\left[\left(\textstyle (\Delta^2 - \gamma^2 - \Delta \xi) + i\gamma(2\Delta - \xi)\right)^{-1/2}\right]  \, s \nonumber\, .
\end{IEEEeqnarray} 
Here, $\deltals(x(t))$ is the light shift derived from Eq.~\eqref{eqn:eigenenergiesTwoLevel} evaluated with the position-dependent detuning $\Delta(x)$, $\Re$ denotes the real part of the expression, and $\xi = E_k/\hbar$ is the oscillation amplitude in units of the detuning. The resulting dispersive curve for a $^{87}$Rb atom with different kinetic energies in a \SI{868}{\nano\meter} lattice ($\chi\approx-0.59$ \cite{Arora2012}) with a trap depth $U_0=k_B\!\times\!\SI{0.5}{\milli\kelvin}$ is shown in the lower panel of Fig.~\ref{fig:model_comparison}(d). 

The most prominent effects on the dispersive curve with increasing temperature are the shift of the inflection point ($\deltals=0$) towards red detuning, the reduction of its amplitude and the increased separation of the extrema. While we observe a clear distortion of the shape for high temperatures it still does not account alone for the observed asymmetric distortion in the experiment. 

The analysis with respect to motional excitations of the atoms, however, gives a hint towards the underlying origin of the asymmetry. The curves in Fig.~\ref{fig:model_comparison}(d) are computed for a fixed position of the atom and identical intensities for all radiation fields. This assumption is no longer valid since in the experiment heating processes occur due to dipole-force fluctuations caused by the repumper and depending on detuning and intensity \cite{Dalibard1985,Taieb1994,Martinez-Dorantes2018} (see Section \ref{sec:imaging} and Appendix \ref{sec:DFFs}). They are especially strong for a red-detuned repumper which for $\Delta<0$ increases the mean kinetic energy of the atoms and shifts the observed $\deltals$ accordingly. As this process is dependent on the repumper detuning, it can explain the observed asymmetric distortion of the dispersive curve. A distinct footprint of this process can be found in the measurement of the survival probability presented in Fig.~\ref{fig:RamanImage}(a). There, the survival drops stronger for $\Delta<0$ indicating a higher mean temperature for this set of experimental parameters.

\section{DIPOLE-FORCE FLUCTUATIONS AND MONTE CARLO SIMULATION}
\label{sec:DFFs}

In the experiment, we observe an asymmetry in the survival probability measurement of Fig.~\ref{fig:RamanImage}(a), which we relate to heating rates that are dependent on the repumper parameters. We attribute those parameter-dependent heating rates to dipole-force fluctuations (DFFs) affecting an atom during the scattering cycles of the repumping process. 
The origin of DFFs in our case is the following: an atom confined in the lattice and illuminated by the repumper field cycles between the ground state $\ket{g}=\ket{5^2 \text{S}_{\nicefrac{1}{2}}}$ and the excited state $\ket{e}=\ket{5^2 \text{P}_{\nicefrac{1}{2}}}$. For $^{87}$Rb at the lattice wavelength of \SI{868}{\nano\meter} the polarizability ratio of the states $\ket{e}$ and $\ket{g}$ is $\chi = \alpha_e / \alpha_g \approx -0.59$, such that the atom experiences a repulsive force for the fraction of time that it spends in the anti-trapping potential of $\ket{e}$, before scattering back into the trapping potential of $\ket{g}$. These fluctuations of the dipole force usually lead to an increase of the kinetic energy of the atom \cite{Dalibard1985,Martinez-Dorantes2018}, although situations leading to cooling have also been observed~\cite{Taieb1994}.

To gain insight onto the effect of DFFs, we implement a semi-classical Monte Carlo simulation of scattering under near-resonant illumination in a 1D optical lattice, following the work in \cite{Martinez-Dorantes2018}. Since the repumper field is close to resonance, we use the dressed states formalism to simulate the motional dynamics of the atom, while taking into account the coherent atom-light coupling. From the simulation we obtain the maps shown in Fig.~\ref{fig:RamanImage}(c) of photon scattering rates and atom loss rates as functions of the repumper detuning and intensity, which validate the hypothesis of DFFs as the main source of heating.

\subsection{Scattering in dressed-state potentials}

Here, we summarize how we treat the simultaneous interaction of the atom with the two relevant light fields: the \SI{868}{\nano\meter} trapping field and the \SI{795}{\nano\meter} repumper field. The interaction of the atom with the lattice field is considered as a pure position dependent AC-Stark shift of the bares states $\ket{g}$ and $\ket{e}$, giving rise to the associated trapping and anti-trapping potentials $U_g(\textbf{r})$ and  $U_e(\textbf{r})=\chi\cdot U_g(\textbf{r})$ respectively, with $|U_g(0)|=U_0$. The additional illumination with the near-resonant repumper field drives scattering transitions between the two atomic states and their respective potentials, generating dynamics with coherent atom-field coupling.

To include the interaction with both light fields in the motional dynamics of the atom, as in \cite{Martinez-Dorantes2018}, we use the dressed-states formalism to describe the evolution of the atom-field system, with scattering events occurring between the repumper dressed states $\ket{-}$ and $\ket{+}$. The associated dressed-state potentials are given by
\begin{equation}
    U_\pm (\textbf{r}) = U_g(\textbf{r}) + \frac{\hbar}{2}\left[-\Delta(\textbf{r}) \pm \sqrt{\Delta^2(\textbf{r}) + \Omega^2} \right], \label{eq:DressedPotentials}
\end{equation}
with $\Omega$ the repumper Rabi frequency on resonance, and $\Delta(\textbf{r})$ the position-dependent detuning taking into account the light-shifts induced by the lattice
\begin{equation}
    \Delta(\textbf{r}) = \Deltat + \frac{U_g(\textbf{r})-U_e(\textbf{r})}{\hbar},
\end{equation}
where $\Deltat$ is the free-space detuning of the light field from the atomic transition. We note that for large values of the detuning $\Delta(\textbf{r})$, the potentials $U_\pm (\textbf{r})$ almost preserve the shape of the bare-state potentials $U_g(\textbf{r})$ and $U_e(\textbf{r})$. When approaching resonance, in the regime where the condition $\Delta(\textbf{r}) =0$ can be met, the shape of the dressed-state potentials and thus the corresponding dipole forces, strongly depend on the parameters of the illumination light, showing the importance of using the dressed states formalism for the simulation.

We follow the approach in \cite{Martinez-Dorantes2018} using the secular approximation of the optical Bloch equations, to obtain the steady-state transition rates between dressed states and their associated potentials \cite{Dalibard1985}
\begin{IEEEeqnarray}{rCl}
	 \Gamma_{++} &=& \Gamma_{--} = \Gamma \sin^2\theta(\textbf{r}) \cos^2\theta(\textbf{r}) , \nonumber \\ 
	 \Gamma_{-+} &=& \Gamma \sin^4\theta(\textbf{r}) , \label{eq:TransitionRates}\\
	 \Gamma_{+-} &=& \Gamma \cos^4\theta(\textbf{r}) , \nonumber
\end{IEEEeqnarray}
with $\theta(\textbf{r})$ the mixing angle defined as
\begin{equation}
    \theta(\textbf{r}) = \frac{1}{2}\arctan\left(-\frac{\Omega}{\Delta(\textbf{r})}\right) + \frac{\pi}{2}H(\Delta(\textbf{r})), \label{eq:MixingAngle}
\end{equation}
where $H(\cdot)$ is the Heaviside step function. 

The Monte Carlo simulation combines the classical motion of a two-level atom in the lattice dressed-state potentials with the semi-classical treatment of scattering induced by the illumination with repumper light. 
We use the position-dependent transition rates of Eq.~\ref{eq:TransitionRates} to calculate and weight the probabilities at a given position that the atom undergoes a scattering event and if it results in a change of potential $U_\pm(\textbf{r}) \rightarrow  U_\mp(\textbf{r})$. Between scattering events (with random sampling), the classical equations of motion in the current potential are solved to determine the position and probabilities of the next scattering event.
Events with change of potential will result in heating by DFFs additional to the photon recoil, while in the case of scattering without state change only the recoil will contribute to the heating dynamics.

\subsection{Details of the Monte Carlo loop algorithm}

The simulation describes the coupled motion and scattering dynamics of a two-level atom (with $^{87}$Rb parameters) trapped in a one-dimensional optical lattice with depth $U_0=k_B\!\times\!\SI{0.5}{\milli\kelvin}$, illuminated with near-resonant repumper light, and considering the dressed-states approach. The polarizability ratio between the two bare atomic states $\ket{g}$ and $\ket{e}$ in the lattice is $\chi=-0.59$. We extract the photon scattering rates (independent of atom losses) and the atom loss rates shown in Fig.~\ref{fig:RamanImage}(c) as functions of the repumper saturation parameter $s=2\left(\Omega/2\gamma\right)^2$ and the free-space detuning $\Deltat$, from the simulated evolution of an ensemble of 500 independent atoms for each set of parameters. We use a linear model to fit the number of total scattered photons per elapsed time, and an exponential model for the increase of kinetic energy from which we determine the atom loss rates.
For each atom with a given set $\{s, \Deltat\}$ we run the following algorithm, starting with the initial conditions:

\begin{itemize}
    \item Initial total energy $E_0$ drawn randomly from a 1D Boltzmann distribution with a mean temperature $T=\SI{100}{\micro\kelvin}$.
    \item Initial position and momentum distributed by one evolution step of the atom in the potential $U_g(r)$, for a random time taken as a fraction of the harmonic oscillator period with trapping frequency $\nu = \SI{350}{\kilo\hertz}$: $\{r=0,p=\sqrt{2m E_0}\}\rightarrow\{r_0,p_0\}$.
    \item Initial state $\ket{-}$ or $\ket{+}$ selected by a weighted coin flip with probabilities $P_{-}=\left\vert\braket{g|-}\right\vert^2 = \cos^2{\theta(r)}$ and $P_{+}=\left\vert\braket{g|+}\right\vert^2 = \sin^2{\theta(r)}$, with $\theta(r)$ the mixing angle from Eq.~\ref{eq:MixingAngle}.
\end{itemize}

And the Monte Carlo loop follows the next steps:
\begin{enumerate}
    \item Start loop iteration $i$ with the atom's position $r_i$, momentum $p_i$, total energy $E_i$ and occupied dressed state determined by the previous iteration (for $i=0$ take the initial conditions).
    \item If in the state $\ket{-}$ ($\ket{+}$), calculate the maximum value of the transition rates $\Gamma_{--}$, $\Gamma_{-+}$ ($\Gamma_{++}$, $\Gamma_{+-}$) within the range of accessible positions at the current energy $E_i$, using Eq.~\ref{eq:TransitionRates}.
    \item Draw random times from the exponential distributions $\rho_\alpha = R_{\alpha,\text{max}} \exp{(-R_{\alpha,\text{max}}t)}$, with $R_{\alpha,\text{max}}$ the maximum values of the two rates from step 2 (for the current dressed state). The minimum of the two resulting times is taken as the time $\tau_i$ elapsed until the next possible scattering event. The transition rate associated to time $\tau_i$ is defined as $R_i(r)$ and it's maximum value $R_{i,\text{max}}$.
    \item Evolve the system a time $\tau_i$ by solving the equations of motion in the dressed-state potential $U_{-}(r)$ ($U_{+}(r)$) in Eq.~\ref{eq:DressedPotentials}, to find $\{r_{i+1},p_{i+1}\}$.
    \item Determine if the scattering event takes place by a weighted coin flip with the success probability $g=R_i(r_{i+1})/R_{i,\text{max}}$. If the event fails to occur, return to step 3.
    \item If the scattering event occurred, update the occupied dressed state accordingly, sum one count to the number of scattered photons, update the new atom's momentum $p_{i+1}$  by adding the photon recoil, and calculate the total energy $E_{i+1}$.
    \item The simulation terminates, if the new position $r_{i+1}$ exceeds the bounds of the trapping potential or if the limit of total time is reached. Otherwise start the next iteration in step 1 with updated values.
\end{enumerate}

\newpage
\bibliography{ramanimaging}

\end{document}